\documentclass[a4paper]{article}
\pdfoutput=1
\usepackage{epsfig}
\usepackage{graphicx}
\usepackage{amsmath}
\usepackage{amssymb}
\usepackage{color}

\usepackage[pagebackref=true,breaklinks=true,colorlinks,bookmarks=false]{hyperref}

\title{On a link between kernel mean maps and Fraunhofer diffraction,
  with an application to super-resolution beyond the diffraction
  limit\footnote{This article has been accepted for publication at the 
    IEEE Conference on Computer Vision and Pattern Recognition (CVPR),
    Portland, 2013.}}

\author{Stefan Harmeling~$^{1}$, Michael Hirsch~$^{1,2}$, Bernhard Sch\"olkopf~$^{1}$\\
$^1$Max Planck Institute for Intelligent Systems, T\"ubingen, Germany\\
$^2$University College London, London, UK\\
\texttt{firstname.lastname@tuebingen.mpg.de}
}

\usepackage{graphicx}
\usepackage{amsmath,amssymb} 
\DeclareMathOperator{\sinc}{sinc}               
\newcommand{\T}{\mathsf{T}}                     
\renewcommand{\H}{\mathsf{H}}                   

\newtheorem{proposition}{Proposition}

\begin{document}

\maketitle

\begin{abstract}
We establish a link between Fourier optics and a recent construction
from the machine learning community termed the kernel mean map. Using
the Fraunhofer approximation, it identifies the kernel with the
squared Fourier transform of the aperture. This allows us to use
results about the invertibility of the kernel mean map to provide a
statement about the invertibility of Fraunhofer diffraction, showing
that imaging processes with arbitrarily small apertures can in
principle be invertible, i.e., do not lose information, provided the
objects to be imaged satisfy a generic condition.  A real world
experiment shows that we can super-resolve beyond the Rayleigh limit.
\end{abstract}

\section{Introduction}

Imaging devices such as telescopes and microscopes collect
incoming light using lenses or mirrors of finite size.  This finite size 
imposes a finite aperture on the light that reaches the optical
system, leading to effects of diffraction. In particular, diffraction ensures that the image of a point can never be a point. For instance, an imaging system using a lens with an $F$-number $f/D$ (where $f$ is the focal length, and $D$ is the diameter of the circular aperture) has an impulse response function (Airy disk) whose radius is $1.22 \lambda f/D$ on the sensor, where $\lambda$ is the wave length of the light (for simplicity, assumed to be monochromatic). 

Another way to express the same insight uses the transfer function. For a lens focused at infinity, the transfer function is constant within a circle of radius $\nu = 1 / (2\lambda f/D)$, and zero outside \cite[p.~136]{SalTei07}. This means, in a nutshell, that if we try to image a sinusoidal pattern with spatial frequency larger than $\nu$, diffraction will annihilate that pattern. Likewise, if we decompose a general object into spatial frequencies by Fourier analysis, all components larger than $\nu$ will vanish.

Similar considerations hold true if, say, an object is scanned by a focused laser beam. Object details smaller than the diffraction limit are washed out, and this fundamental limit of image-formation systems is often referred to as the \emph{diffraction limit} \cite[p.~136]{SalTei07}. There are ways to circumvent it using sophisticated hardware, for instance with scanning near-field optical microscopy, or stimulated emission depletion microscopy (STED) using fluorescence \cite{hell1994breaking}, but these are not the topic of the current paper. Instead, we want to assay whether restrictions on the object being imaged can fundamentally change the resolution of an optical system. Specifically, we will show that under the generic assumption of bounded support, one can in principle (i.e., given a perfect measurement of the image) resolve arbitrarily fine detail. This is done by pointing out a connection to the field of kernel methods in machine learning, and utilizing certain theoretical results from that domain.
We do not claim that all our insights are new --- indeed, we will point out that in spite of the above received wisdom, there are certain theoretical results in the optics community, some of them rather old, that draw similar conclusions. We do believe, however, that the link to kernel methods is new, and hope that it will lead to a fruitful cross-fertilization of two previously unconnected branches of research.
Using toy examples, we show that the assumption of bounded support can be used to recover image detail past the diffraction limit for simple real-world images, which are pixelized and not noise-free.

The paper is structured as follows. In Section~\ref{kernelmeans}, we
explain the notion of kernel means. These are particular types of
mappings into reproducing kernel Hilbert spaces, and in some cases
they can be shown to be invertible.  The kernel map has applications
in a number of tasks including testing of homogeneity and independence
\cite{GreBorRasSchetal07,GreFukTeoSonetal08}.  However, our main
interest is a link to wave optics, to be described in the next
section.\footnote{This link was pointed out during a
  mathematical workshop in Oberwolfach, see \cite{Sch2008}.}  In Section~\ref{fraunhofer}, we explain some basics of
Fourier optics, in particular the Fraunhofer approximation of
diffraction. We show that Fraunhofer diffraction is actually a
particular case of kernel mean mapping. This link between Fourier
optics and machine learning allows us to leverage some theoretical
results about kernel mean maps to make a surprising statement about
super-resolved imaging. Section~\ref{related} discusses how this
result relates to certain observations made by the wave optics
community.

\section{Characteristic kernel means}
\label{kernelmeans}
A symmetric function $k:{\mathcal X}^2\to {\mathbb{R}}$, where
${\mathcal X}$ is a nonempty set, is called a \emph{positive definite (pd) kernel} if for arbitrary points
$x_1,\dots,x_m\in {\mathcal X}$ and
coefficients $a_1,\dots,a_m\in {\mathbb R}$, we have
$$\sum_{i,j}a_ia_jk(x_i,x_j)\ge 0.$$
The kernel is called \emph{strictly positive definite} if moreover for pairwise
distinct points equality with zero, $\sum_{i,j}a_ia_j k(x_i,x_j)=0$,
implies that all coefficients vanish, $a_i=0$ for all $i$.

Any positive definite kernel induces a mapping 
\begin{equation}\label{eq:featuremap}
x\mapsto k(x,.)
\end{equation}
into a reproducing kernel Hilbert space (RKHS), which
is a Hilbert space of functions $f:{\mathcal X}\rightarrow {\mathbb R}$
with an inner product $\langle .,.\rangle$ such that $k$ represents point evaluation,
\begin{align}
  \langle f(.),k(x,.)\rangle = f(x)
\end{align}
which implies also the \emph{reproducing property} $\langle
k(x,.),k(x',.)\rangle = k(x,x')$, see e.g.~\cite{SchSmo02} for more details.

\subsection{Kernel mean of a sample}
In an SVM \cite{SchSmo02}, (\ref{eq:featuremap}) is the mapping that takes each datapoint into the so-called feature space, in which a linear learning method is applied. Rather than mapping the points one by one, however, one can also map a sample or a distribution directly to its mean in the feature space. Below, we will show that this kind of mapping contains optical imaging as a special case. But before, we first point out that even though the operation of taking the mean usually comes with a loss of information, this need not be the case if the kernel satisfies a certain condition.

Consider a sample of points $X= \{ x_1,\dots,x_m\}\subset {\mathcal X}$, that are distinct, i.e.,\ $x_i\neq x_j$ whenever $i\neq j$.
Given a pd kernel $k$, we define the \emph{kernel mean map} of $X$ by \cite{SchSmo02,smola2007hilbert}
\begin{equation}
\mu(X) = \frac{1}{m}\sum_{i=1}^m k(x_i,\cdot).
\end{equation}

Consider another sample of distinct points $Y= \{ y_1,\dots,y_n\}\subset {\mathcal X}$.
Clearly, if $X$ equals $Y$, their kernel means are
identical.  What about the converse?

We call a kernel \emph{characteristic for samples}, if the mean map
$\mu$ based on $k$ is injective, i.e., if identical kernel means $\mu(X)=\mu(Y)$ imply
identical samples $X=Y$.

It is not obvious whether characteristic kernels exist.  E.g.~for
polynomial kernels $k(x,x')=(\langle x, x'\rangle+1)^d$, with $d\in{\mathbb N}$, observing equal kernel means $\mu(X)=\mu(Y)$ for the samples $X$ and $Y$ implies that all empirical moments up to order $d$ of $X$ and $Y$
coincide.  However, $X$ and $Y$ might differ in their empirical
moments of higher orders.  The following proposition gives a
sufficient condition for being a characteristic kernel:
\begin{proposition}
Strictly pd kernels are characteristic for samples.
\end{proposition}
\textit{Proof:} Consider a strictly pd kernel $k$ and its
mean map $\mu$.  Consider two samples $X=\{ x_1,\dots,x_m\}\subset {\mathcal X}$ and $Y= \{ y_1,\dots,y_n\}\subset {\mathcal X}$ as above with equal
kernel means, $\mu(X)=\mu(Y)$.
Let $Z=\{z_1, \ldots, z_l\}$ be the set (not the multiset) of all
elements in the union of $X$ and $Y$, i.e.~all elements in $Z$ are
pairwise distinct.  Let $\#X(z)$ be the number of times $z$ appears in
$X$, similarly $\#Y(z)$.  Define $\gamma_i=\#X(z_i)/m - \#Y(z_i)/n$.
Then we have 
\begin{align}\label{eq:lincombs}
0&=\mu(X)-\mu(Y)\\
&=\sum_{i=1}^m \frac{1}{m} k(x_i,.) - \sum_{i=1}^n \frac{1}{n} k(y_i,.)
=\sum_{i=1}^l \gamma_i k(z_i,.)\label{eq:zerovector}
\end{align}
Now take the dot product between (\ref{eq:zerovector}) and itself, leading to
\begin{equation}
0 =  \langle \sum_{i=1}^l \gamma_i k(z_i,.), \sum_{j=1}^l \gamma_j k(z_j,.) \rangle,
\end{equation}
which by the reproducing property and bilinearity amounts to
\begin{equation}
0 = \sum_{i,j=1}^l   \gamma_i\gamma_j k(z_i,z_j).
\end{equation}
Since $k$ is strictly pd, this implies that for all $i$ the coefficients
$\gamma_i$ are zero, thus $\#X(z_i) = \#Y(z_i) m/n$.
Since $\#X(z_i), \#Y(z_i) \in \{0,1\}$, we conclude that $m=n$ and $\#X(z_i) = \#Y(z_i)$ for all $i$, i.e., $X=Y$.

\hfill\rule[-0.4mm]{2.0mm}{3.2mm}

The mean map has some other interesting properties \cite{smola2007hilbert}. Among them is the
fact that $\mu(X)$ represents the operation of taking a mean of a
function on the sample $X$:
\begin{align}
  \langle \mu(X),f \rangle = \left\langle \frac{1}{m}\sum_{i=1}^m
    k(x_i,\cdot),f \right\rangle = \frac{1}{m}\sum_{i=1}^m f(x_i)
\end{align}
where we have applied the point evaluation property.

\subsection{Kernel mean of a probability measure}

Instead of samples we next consider probability measures\footnote{We
  assume that all measures considered are Borel measures.} defined on
$\mathcal X$ assuming that $\mathcal X$ has the necessary additional
structure.  To ensure that the following integrals exists, we assume
that all considered kernels are bounded (see
\cite{sripermbudur2010}). Below, we will think of the measures as the light distribution of the object being imaged.
 We extend the mean map to
probability measures by defining the \emph{kernel mean} of $P$ as
\begin{align}
  \mu(P) = \int k(x,.) \;dP(x).
\end{align}
Similar to the above definition, we call a kernel
\emph{characteristic for probability measures} \cite{FukGreSunSch08} if the mean map is
injective for probability measures, i.e., $\mu(P)=\mu(Q)$ implies that
$P$ and $Q$ are equal.

To state the analog of Proposition 1, we define a kernel $k$ to be
\emph{integrally strictly positive definite} if for any finite non-zero
signed Borel measure $\nu$, the integral of $k$ wrt. $\nu$ is strictly positive,
\begin{align}
  \int k(x, x') \;d\nu(x) \;d\nu(x') > 0.
\end{align}
Note that an integrally strictly pd kernel is also strictly pd but not
vice versa.
\begin{proposition}
Integrally strictly pd kernels are characteristic for
probability measures.
\end{proposition}
This result was proven by \cite{sripermbudur2010}; we only provide a brief
\textit{proof sketch:} Consider two different probability measures $P$ and
$Q$.  Their
difference is a finite non-zero signed Borel measure $\nu=P-Q$.  Assuming equal kernel
means, we have:
\begin{align}
  0 &= \mu(P)-\mu(Q)\\
  &=  \int k(x,.) \;dP(x) - \int k(x,.) \;dQ(x)\\
  &= \int k(x,.)\;d\nu(x)
\end{align}
Taking the squared norm and using the reproducing property we get a
contradiction,
\begin{align}
  0 &= \langle\int k(x,.)\;d\nu(x), \int k(x,.) \;d\nu(x)\rangle \\
  &= \int k(x,x') \;d\nu(x)\;d\nu(x')> 0
\end{align}
where we used for the last inequality the fact that $k$ is integrally
strictly pd.  
\hfill\rule[-0.4mm]{2.0mm}{3.2mm}

A more specific view on characteristic kernels, which will apply in
the case of Fraunhofer imaging, can be obtained by considering
\emph{translation invariant} pd kernels on ${\mathcal X}={\mathbb
  R}^d$, i.e., kernels that can be written as $k(x,x')=\psi(x-x')$
with some continuous function $\psi:{\mathbb R}^d \rightarrow {\mathbb
  R}$. By Bochner's theorem \cite{wendland2005scattered}, they can be
expressed as the Fourier transform of a finite non-negative Borel
measure $\Lambda$,
\begin{align}
  \psi(x) = \int e^{-ix^\T\omega}\;d\Lambda(\omega).
\end{align}
Following Corollary 4 in \cite{sripermbudur2010} we can write the
squared RKHS distance between the kernel means of two probability measures in terms of their
characteristic functions,
\begin{align}
  \| \mu(P)-\mu(Q) \|^2 
  &= \int |\phi_P(\omega)-\phi_Q(\omega)|^2\; d\Lambda(\omega)
\end{align}
where $\|.\|$ is the norm of the RKHS and $\phi_P(\omega)=\int
e^{ix^\T \omega}\;dP(x)$ is the characteristic function of $P$,
and likewise $\phi_Q$.  Roughly speaking, this shows that $P$ and $Q$ can
be distinguished as long as the spectrum $\Lambda$ of the kernel is
nonzero wherever the spectra of the probability distributions might
differ. If $\Lambda$ has full support, i.e.~it is non-zero almost
everywhere, the corresponding kernel can distinguish all probability
distributions.  If it does not have full support, it can sometimes still
distinguish a restricted class of probability distribution as we see
next.

\subsection{Kernel mean of a probability measure with bounded support}

Consider a translation invariant pd kernel $k$ such that the support
of the corresponding $\Lambda$ has a non-empty interior.  For what
class of probability measures can such a kernel be
characteristic\footnote{We use \emph{characteristic for a class of
    probability measures} in the obvious way, i.e.~the kernel map is
  injective for the restricted class.}?
An obvious choice is a class of probability measures whose
characteristic functions agree outside the support of $\Lambda$.
However, there is a much more interesting class of measures which we
define next.

Let us consider a probability measure $P$ with compact support.  By the
Paley-Wiener theorem \cite{rudin91} its characteristic function
$\phi_P$ is entire (aka analytic or holomorphic), which implies that
knowing $\phi_P$ on a compact subset determines $\phi_P$ everywhere.
This leads to the following proposition:
\begin{proposition}\label{boundedprop}
  Translation invariant pd kernels, whose corresponding $\Lambda$
  have a support with non-empty interior, are characteristic for
  probability measures with compact support.
\end{proposition}
This is a simplification of Theorem 12 in \cite{sripermbudur2010}
which also contains a detailed proof.

The kernel which will be relevant in the next section is the sinc
kernel defined for $\sigma>0$ as
\begin{align}
  k(x,x') = \psi(x-x') = \frac{\sin \sigma (x-x')}{x-x'}.
\end{align}
The Fourier transform of $\psi$ is the scaled indicator function of
the interval $[-\sigma,\sigma]$, i.e.
\begin{align}
  \Lambda(\omega) = \sqrt{\frac{\pi}{2}}\; 1_{[-\sigma,\sigma]}(\omega),
\end{align}
so $\Lambda$ is non-zero on that interval (thus having a support with
non-empty interior) and is thus
characteristic for probability measures of bounded support.  The
square of the sinc kernel has the same properties, since it
corresponds to the convolution of $\Lambda$ with itself, inheriting a support with non-empty interior from $\Lambda$.

\section{Incoherent imaging as a mean map}
\label{fraunhofer}

\subsection{Imaging under incoherent illumination}
As electromagnetic radiation, light is governed by Maxwell’s equations
–-- a set of linear partial differential equations that form the
foundation of classical electrodynamics including classical
optics. Although electric and magnetic fields are vectorial in nature,
in many situations\footnote{More precisely, the scalar theory of
  electromagnetism is valid in linear, isotropic, homogeneous and non-dispersive dielectric media such as free space or a lens with
  constant refractive index, where all components of the electric and
  magnetic field behave identically} polarisation effects, i.e. any
coupling between the electric and magnetic fields, can be neglected
and all components of the electric and magnetic field can be well
described by a single scalar wave equation \cite{Hirsch11}
\begin{align}
(\nabla^2-\frac{n_0^2}{c^2}\frac{\partial^2}{\partial t^2}) \, \Phi(u,t) = 0,
\label{eq:scalarwave}
\end{align}
where $\Phi(u,t)$ is any of the scalar field components of the
electric or magnetic field and $n_0$ denotes the refractive index of the
medium, within which the light is propagating. Since
(\ref{eq:scalarwave}) is a \emph{linear} partial differential
equation, any linear combination of its solutions yields another
solution. The property of linearity has major implications for the
mathematical treatment as it allows us to analyse a system by studying
its response to a single point stimulus. Its effect to a complex input
signal $\Phi(\xi,t)$ can be obtained by considering the input signal
being composed of point stimuli and adding up their known responses
accordingly:
\begin{align}
  \Psi(u,t) = \int h(u-\xi) \,\, \Phi(\xi,t) \; d\xi.
  \label{eq:incoherent2}
\end{align}
Here $\Psi$ denotes the output of a \emph{linear optical system} which
is fully described by its impulse response $h(u-\xi)$. For ease of
exposition we implicitly assume stationarity both in space
(i.e. $h(u;\xi)=h(u-\xi)$) and time (i.e. $h$ depends not on $t$) in
(\ref{eq:incoherent2}).

Optical detectors such as CCD sensors usually record intensities,
i.e. the square of the field amplitude. Since the integration time is
much longer than a single period of oscillation, we must average over
time to obtain the recorded pixel intensities
\begin{align}
\left \langle \Psi(u,t) \bar{\Psi}(u,t)  \right \rangle
= \iint h(u-\xi) \, \bar{h}(u-\xi') \,\times \\ \left
  \langle \Phi(\xi,t)\,\bar{\Phi}(\xi',t)\right \rangle \; d\xi\,
d\xi',
\label{eq:imagingequation}
\end{align}
where $\left \langle .  \right \rangle$ denotes temporal averaging.
Here, we must take the coherence properties of the light into
account and distinguish between \emph{coherent} and \emph{incoherent
  illumination}:
\begin{itemize}
\item In the case of \emph{coherent illumination}, we cannot simplify
  Equation (\ref{eq:imagingequation}) any further without making any
  additional assumptions. The square of the complex field can lead to
  cancellations or other non-linear \emph{interference} effects.
\item In the case of \emph{incoherent illumination}, the spatial
  correlation between any two light rays emitted from the scene is
  assumed to be negligible. Hence, the time average in
  (\ref{eq:imagingequation}) will only contribute to the integral for
  $\xi = \xi'$:
  \begin{align}
    \left \langle \Phi(\xi,t)\,\bar{\Phi}(\xi',t)\right \rangle &=
    |\Phi(\xi)|^2 \,\, \delta(\xi-\xi')
 \label{eq:inco}
  \end{align} 
\end{itemize}
Plugging expression (\ref{eq:inco}) into Equation
(\ref{eq:imagingequation}) yields the \emph{incoherent imaging
  equation} 
\begin{align}
  q(u) = \int f(u-\xi) \,\, p(\xi) \; d\xi,
  \label{eq:incoherent}
\end{align}
where we introduced $q(u)$, $p(\xi)$ and $f(u-\xi)$ for $\left\langle
  |\Phi(u,t)|^2 \right \rangle$, $\left\langle|\Psi(\xi,t)|^2\right
\rangle$ and $|h(u-\xi)|^2$, respectively. Both $p(\xi)$ and $q(u)$
describe image intensities; the impulse response $f$ is called the
\emph{point spread function} (PSF) of the imaging system as it
corresponds to the image of a point light source. 

Although we had to make a number of assumptions to derive the
incoherent imaging equation (\ref{eq:incoherent}), it has been found
to provide an accurate description for most typical imaging systems
including astronomical, microscopical imaging and photography
\cite{Barnes_1971}.

\subsection{Connection to kernel mean map}
As an image is inherently non-negative, the image of the object
$p(\xi)$ induces, up to normalization, a probability measure $P$. In
addition we assume finite energy, i.e., $\int p(\xi) d\xi<\infty$. Then
Eq.~(\ref{eq:incoherent}) can be understood such that such that for
the translation-invariant kernel function $k(u,\xi)=f(u-\xi)$,
the resulting image $q$ is the kernel mean of
$P$:
\begin{align}
  \label{eq:1}
  \mu(P) = q(.)
\end{align}
So we obtained the interesting result that the incoherent imaging
equation can be expressed as a kernel mean.\footnote{This provides a
  physical interpretation of the kernel as the point response of an
  optical system. This kind of interpretation can be beneficial also
  for other systems, and indeed it is suggested by the view of kernels
  as Green's functions \cite{HofSchSmo08,SchSmo02}: the kernel $k$ can
  be viewed as the Green's function of $P^*P$, where $P$ is a
  regularization operator such that the RKHS norm can be written as
  $\|f\|_k = \|P f\|$. For instance, the Gaussian kernel corresponds
  to a regularization operator which computes an infinite series of
  derivatives of $f$. }

\subsection{Fraunhofer diffraction}
The resolution of any optical system even without optical aberrations
is limited by diffraction. The mathematical framework describing
diffraction is Fourier optics \cite[e.g.]{SalTei07}. It decomposes the
light radiated by an object into harmonic components of different
spatial frequencies, each one corresponding to a plane wave whose
amplitude is given by the Fourier transform of the emitted light
field. It turns out that at a far distance from the object, most of
these waves cancel each other, and each direction in space only 'sees'
one of the plane waves --- the free-space wave propagation can be
identified with the Fourier transform, different spatial frequencies
in the object corresponding to one direction each. This is referred to
as the \emph{Fraunhofer approximation}. By means of a lens, this
situation can be realised also for a finite distance, and different
directions in space correspond to different coordinates on the image
plane, or camera sensor.

In an ideal, aberration-free optical system, the Fraunhofer
approximation states that the PSF is the inverse Fourier transform of
the auto-correlation function of the pupil or aperture function
\cite{goodman96}. In the following we compute the PSF for the simple
case of a circular planar aperture.

\subsection{Diffraction in one dimension}

In one dimension, consider an aperture
$a:{\mathbb R}\rightarrow{\mathbb R}$ 
defined as $a(\omega) = 1_{[-\sigma, \sigma]}(\omega)$.  The inverse
Fourier transform of $a$ is the sinc function $\sin(\omega x)/x$.
Then by the Wiener-Khinchin theorem the PSF $f$ as the
auto-correlation function of the aperture function, i.e. $a$, is the
square of the $\sinc$ function,
\begin{align}
  f(x) = \left(\frac{\sin(\omega x)}{x}\right)^2.
\end{align}

\subsection{Diffraction in two dimensions}

Also for more than one dimension the incoherent imaging equation is
expressible as a kernel mean.  For this we consider a two dimensional
circular aperture with radius $\sigma$, where the aperture function is
the pill box function:
\begin{align}
  a(\omega) = \left\{\begin{array}{cl} 1 & \text{if } \|\omega\|\le \sigma\\ 0 & \text{otherwise}
    \end{array}\right.
\end{align}
Again, the PSF is the Fourier transform of the auto-correlation
function, which in this case is the squared Bessel function of the
first kind of order one,
\begin{align}
  f(x) = \left(\frac{J_1(\omega x)}{x}\right)^2.
\end{align}
Note that any translation-invariant kernel $k$ constructed from a
positive aperture function is pd due to Bochner's theorem, so the
corresponding diffraction can be written as a kernel mean as in Eq.~(\ref{eq:1}).
Note that in addition to the two apertures discussed so far, we could use arbitrary apertures satisfying the condition of Proposition~\ref{boundedprop}, including apertures that are not indicator functions (if physically realizable): Bochner's theorem ensures that for all nonnegative measures, the Fourier transform is a pd kernel, and Proposition~\ref{boundedprop} ensures that the kernels are characteristic.

\subsection{Breaking the diffraction limit}

The actual resolution that is possible with a given optical system is
determined by the size of the aperture, which could be the size of the
mirror or lens in a telescope.  

Having written the incoherent imaging equation as kernel means, we can
apply the insight from the previous section to obtain the surprising
result that an object $p(\xi)$ with bounded support, i.e.~$p(\xi)$ is zero
outside some compact area, the Fraunhofer diffraction does not destroy
any information, i.e.~at least theoretically, the diffraction limit
is no limit:
\begin{proposition}
  An object with bounded support can be recovered completely
  from its diffraction-limited image.
\end{proposition}
\textit{Proof:} This follows from the injectivity of $\mu$ in
the context of Proposition \ref{boundedprop} and the fact that any aperture shape induces a translation-invariant pd kernel by Bochner's theorem.

Note that this proposition only states that the kernel mean map is
invertible --- it does not make a statement about the practical
problem of how to compute the inverse.  In the next section we
present a simple approach to do so.

\section{Experiments}
\begin{figure}[t]
  \centering
  \includegraphics[width=\columnwidth]{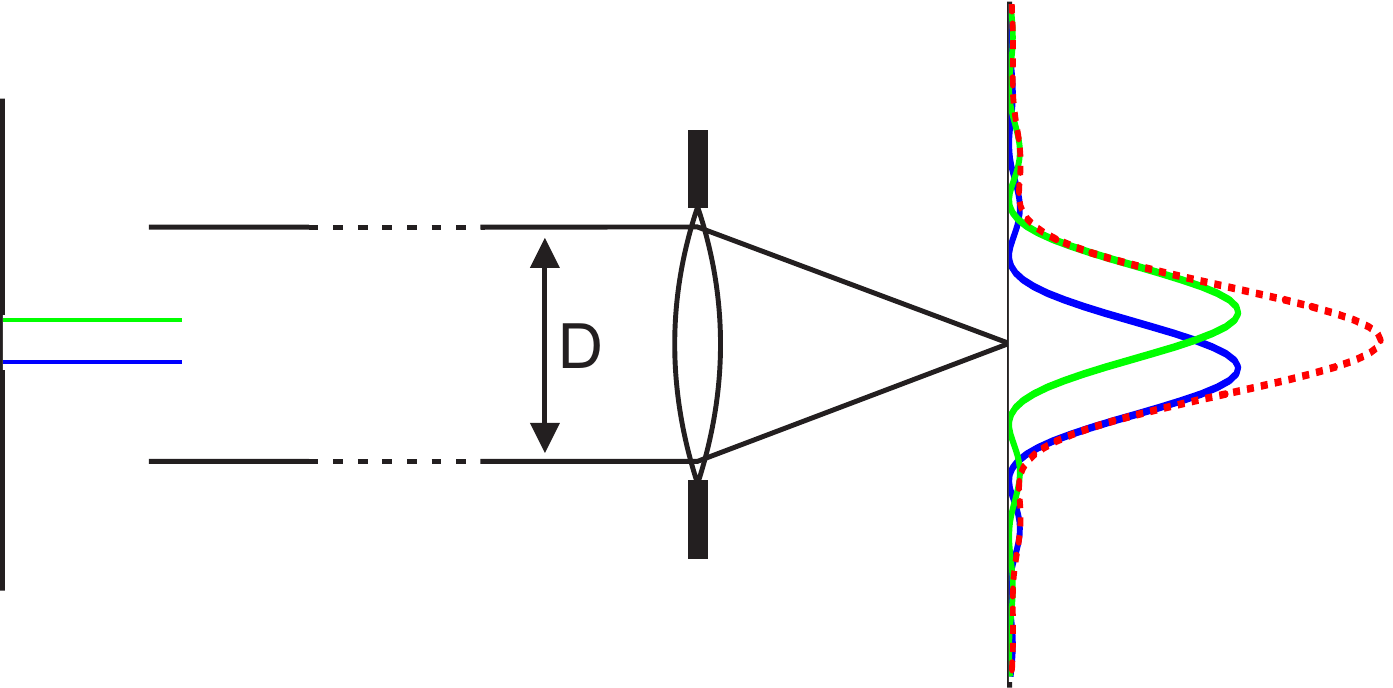}
  \caption{A one dimensional double star (two delta peaks on the left) gets imaged by the lens with the finite aperture leading to an blurred image formed by the sum of two squared sinc functions on the right.}
  \label{fig:schema}
\end{figure}

Fig.~\ref{fig:schema} illustrates a typical experimental setup: two
point sources (in green and blue on the left) are imaged through an
optical system consisting here of a single lens (with focal length $f$) and a finite aperture of
diameter $D$. Under incoherent illumination the observed image on the
right is a superposition of the images of the point sources, each of
which is given by the impulse response of the optical system
$\Psi$. In an ideal diffraction-limited optical system, two point
sources can only be resolved if they are at least $1.22\lambda f/D$
apart. To demonstrate that we can resolve beyond this so-called Rayleigh limit,
we place the two point sources so close, that their individual images cannot be
resolved (i.e.~the red dashed line in Fig.~\ref{fig:schema} has only
one maximum).  

\subsection{Recovering a one-dimensional simulated image}

\begin{figure}
  \centering
  \includegraphics[width=\columnwidth]{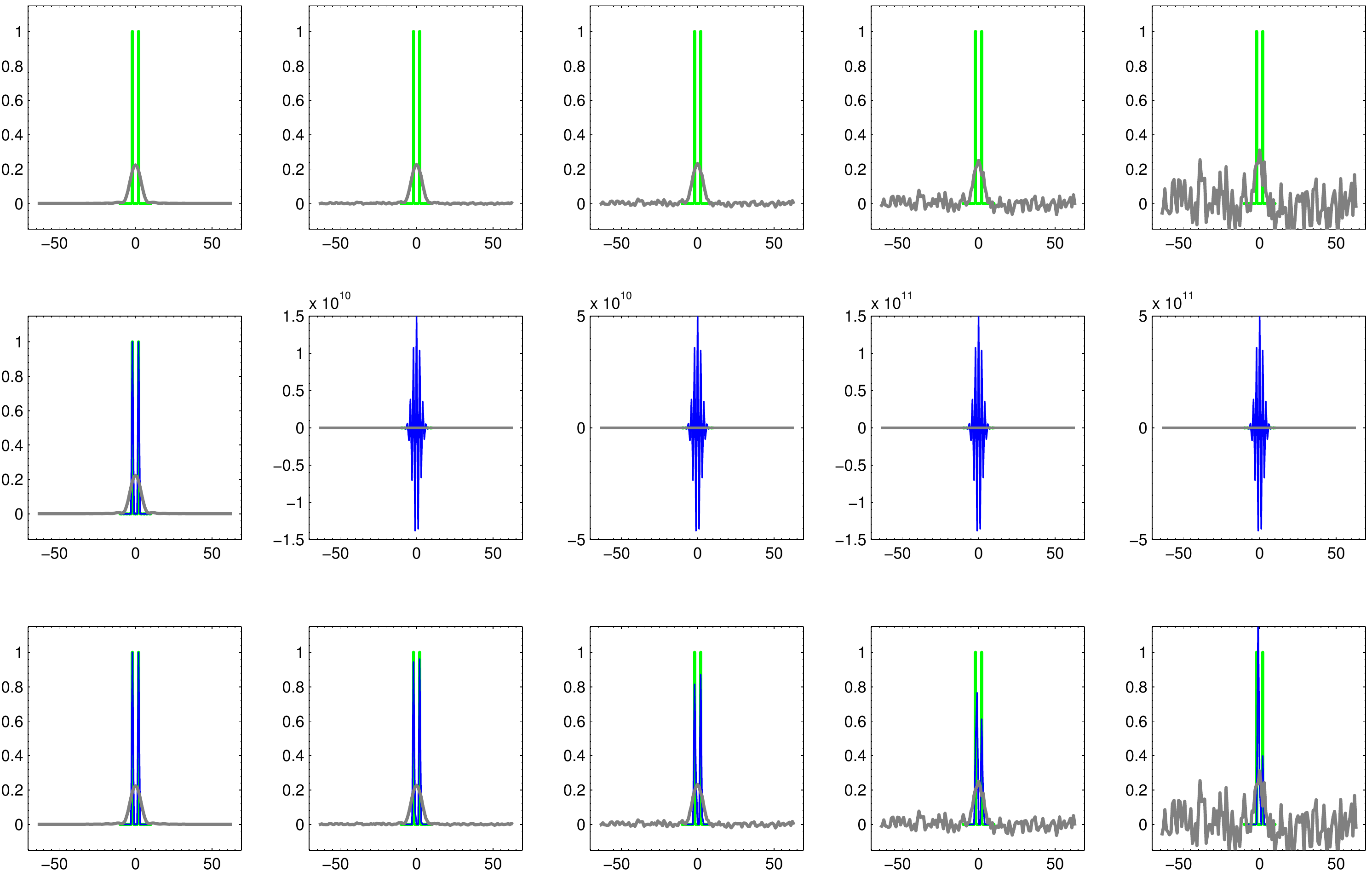}
  \caption{Restoring a diffraction-limited image (gray, first row) of one-dimensional double star (green, first row) with increasing amounts of noise (from left to right).  The maximum likelihood solution (blue, second row) restores the double stars only in the noise-free cases (left column).  The non-negatively constrained maximum likelihood approach (blue, third row) restores the double star even for various amounts of noise (third row, left to right).}
  \label{fig:exp}
\end{figure}
The recorded image is usually corrupted by
measurement noise, sometimes modeled as additive Gaussian.
Then Eq.~(\ref{eq:1}) becomes
$q(.) = \mu(P) + n$
where $n \propto N(0,\sigma)$. The first row of Fig.~\ref{fig:exp}
shows the true object (green) and the observed image (gray) of a one dimensional toy example
for increasing amounts of noise (from left to right). More precisely, we represent the true
object $p$ and the recorded image $q$ as finite-length one-dimensional column vectors $u$ and $v$.  According to the
Fraunhofer diffraction equation, the relationship between the object $u$ and image $v$ is linear and can be expressed as a matrix:
\begin{align}
  \label{eq:11}
  v = F^\H  T  F  Z  u + n.
\end{align}
Here, $Z$ is a zero-padding matrix, $F$ is the discrete Fourier
transform matrix, $F^\H$ the hermitian matrix of $F$ (i.e.~the inverse transform), and $T$ is the optical transfer
function (OTF), i.e.~the Fourier transform of the system's impulse
response, i.e.~$T=F\psi$, with $\psi$ being a finite dimensional vector, too.

The object $u$ can be recovered from $v$ by a maximum likelihood approach, i.e.~we solve 
the following least-squares problem
\begin{align}
  \label{eq:12}
  \text{min}_u \| v - F^\H  T  F  Z  u \|^2.
\end{align}
The middle row of Fig.~\ref{fig:exp} shows the recovered objects
$u$ of the noisy observations $v$ (first row in gray) using the Matlab
command 
\begin{verbatim}    
                       u = (F'*T*F*Z) \ v;
\end{verbatim}
As suggested by  our
findings in Section \ref{fraunhofer}, the true signal can  be recovered
exactly in the noise-free case (first column). The assumption of
bounded support is implicit by chosing $u$ to be shorter than $v$. However, already small amounts of noise render the
optimisation problem in Eq.~(\ref{eq:12}) ill-conditioned yielding an
unstable solution.

As an image accounts for the amount of recorded photons we can employ
non-negativity as an additional physical constraint. Hence, instead of
Eq.~(\ref{eq:12}) we solve the constrained optimization problem   
\begin{align}
  \label{eq:13}
  \text{min}_u \| v - F^\H  T  F  Z  u \|^2 \text{ s.t. } u \geq 0.
\end{align}
The non-negativity constraint stabilizes the restoration process and
yields good results even for large amounts of noise (bottom row in
Fig.~\ref{fig:exp}).  We solve the non-negative least squares problem
using the Matlab command:
\begin{verbatim}
                   u = lsqnonneg(F'*T*F*Z, v);
\end{verbatim}
\subsection{Recovering a two-dimensional real image}

\begin{figure*}
  \centering
  \begin{tabular}{cc}
   \includegraphics[width=0.45\textwidth]{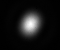}
  &\includegraphics[width=0.45\textwidth]{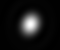}\\
  (a) $v$, aperture=0.5mm
  & (b) $D^\H TDZu$\\[5mm]
   \includegraphics[width=0.45\textwidth]{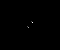}
  &\includegraphics[width=0.45\textwidth]{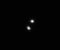}\\
  (c) $u$, recovered image
  & (d) ground truth, aperture=4mm
  \end{tabular}
  \caption{Real photograph of an artifical double star, that is
    clearly visible if the aperture is open (d), but not for small
    aperture (a).  The recovered image (c) shows the two stars without
  blur, (b) shows the result of passing (c) through the forward
  model.  All images show crops (size $60\times 50$) of larger
  images (size $647\times 570$).}
  \label{fig:real}
\end{figure*}

We build an experimental setup with an artificial double star (lighted
by green light) that is imaged by a cooled camera (PCO.2000) in about one
meter distance.  The optics of the camera consists of a changeable
aperture and a single lens ($f=100$mm).  Panel (d) of
Fig.~\ref{fig:real} shows a ``ground truth'' image that has been taken
with an aperture of $4$mm and exposure time of $3$ms.  Panel (a) shows
the same double star but with aperture $0.5$mm.  The aperture has been
chosen that the angular separation of the double star is 50 percent
below the Rayleigh limit.
Note that the two
stars are not visible anymore and the light has been spread out due to
diffraction.  To get a good measurement we had to expose for $4000$ms.
Both images, (a) and (d), are the result of averaging eight images
minus an averaged dark frame to reduce the noise to a minimum.  The
support is chosen by thresholding the measured image, panel (a).
Applying the method described in the previous paragraph to the image
in panel (a), we are able to recover the two double stars which are
quite similar to the ground truth (panels (c) and (d) in
Fig.~\ref{fig:real}).  Note that the ground truth is more blurry since
it is also photographed with a finite aperture.  

\section{Related work}
\label{related}

The question whether it is possible to break the diffraction limit has been the
subject of numerous works:

In 1952, Toraldo di Francia \cite{toraldo52} stated that ``we
notice that the classical limit of $1.22 \lambda/D$, which has always
been accepted as a \emph{theoretical} limit, proves instead to be only
a \emph{practical} limit.''  Motivated
by ``super-gain antennas'' he studies the diffraction patterns of
``super-resolving pupils'' which consists of concentric rings instead
of a uniform pupil.  He observes that for an increasing number of
rings the central disc of the airy disc becomes smaller and more
isolated, hereby increasing the resolution.
In \cite{toraldo55}, the same author discusses the problem of
resolving power from the point of view of information theory.  He
makes the point that several objects can lead to the same image, so
without an ``infinite'' amount of prior information we cannot do
two-point resolution.

A few years later, Wolter showed in \cite{wolter59b} that bounded
illumination (cf.\ our bounded support assumption on the object), is
sufficient to recover higher frequencies, since the Fourier transform of a
bounded object is analytic.  He uses accelerating summation techniques
to analytically continue the spectrum that has been cut off by an
aperture.
Independently of Wolter, Harris \cite{harris64} also considered bounded
objects and the fact that their Fourier transforms are analytic.  He
also proposed a method for analytic continuation (for the noise-free
case).  His conclusion is that 
``diffraction imposes a resolution limit which is determined by the
noise of the system rather than by some absolute criterion.''

Barnes \cite{barnes66} proposed a reconstruction procedure for
coherent illumination.  He uses the assumption of bounded support to
write the convolution operator in the imaging equation in such a way
that it can be decomposed into prolate spheroidal wave functions
\cite{slepian61}.  This allows inversion of that operator, similar to
division in Fourier space.
Rushforth and Harris \cite{rushforth68} study the influence of noise
on reconstruction methods to overcome the diffraction limit.  Their
conclusion is that ``the Rayleigh criterion is an approximate measure
of the resolution which can be achieved easily.''

Gerchberg \cite{gerchberg74} (and independently Papoulis
\cite{papoulis75}) proposed an algorithm analogous to Gerchberg and
Saxton's phase retrieval method \cite{gerchberg1972practical}
incorporating also positivity.  As Jones \cite{jones1986discrete}
points out, this algorithms converges under certain conditions only rather slowly.

Although the above works have provided insight into theoretical aspects of recovering object properties
beyond the diffraction limit, the proposed methods did not become relevant in practice. 
In 1993, Sementilli, Hunt and
Nadar \cite{sementilli93} derived bounds on the bandwidth extension in
terms of object size and noise variance under the assumption of
bounded object support and positivity.
Section 6.6 of Goodman's book on Fourier Optics \cite{goodman96}
discusses these early studies of the diffraction limits and concludes,
that ``the Rayleigh limit to resolution represents a practical limit
to the resolution that can be achieved with a conventional imaging
system.'' 

Several papers consider a bounded support
constraint to overcome the diffraction limit.  Another possible
constraint is sparsity:  Donoho \cite{donoho91} studied the problem of
recovering a sparse signal for which only low frequencies of its
Fourier transform are available.  Recently, Candes and
Fernandez-Granda \cite{candes2012towards} also studied conditions under which
sparse signals can
be recovered.  The results apply to signals which have a
sparse representation. Sparsity has effectively also been practically used to break the diffraction
limit using hardware, e.g.~in stimulated emission
depletion microscopy (STED) \cite{hell1994breaking}.

Finally, one should mention that the works above consider
superresolution as the problem of breaking the diffraction limit, as opposed to trying to ``only'' increase the
resolution of low resolution sensors (e.g.~\cite{irani90}).  This type
of superresolution is not the topic of this paper so we refer the
reader to the review of Park, Park and Kang
\cite{park2003super}.

\section{Conclusion}

We have developed a novel connection between machine learning and
Fourier optics, identifying a positive definite kernel with the
squared Fourier transform of an imaging system's aperture. Leveraging
results from RKHS theory, this led to a condition on an object
(boundedness of its support) which ensures that it can be
fully reconstructed from the image. Simple experiments showed that such reconstructions are possible with real data. While we do not claim that our approach has immediate practical implications, we believe it is surprising and noteworthy that a celebrated results in Fourier optics can be analyzed using the theory of positive definite kernels used in machine learning, with nontrivial implications for the profound problem of optical super-resolution.
We hope this link can be further exploited to
gain a beter understanding and possibly novel solutions to optical
problems.  In an experimental setup we show that we are able to
super-resolve beyond the Rayleigh limit.

{\small
\bibliographystyle{plain}
\bibliography{beyond_abbe_arxiv}

\begin{thebibliography}{10}

\bibitem{barnes66}
C.W. Barnes.
\newblock Object restoration in a diffraction-limited imaging system.
\newblock {\em Journal of the Optical Society of America}, 56(5):575--578, May
  1966.

\bibitem{Barnes_1971}
K.R. Barnes.
\newblock {\em {The Optical Transfer Function}}.
\newblock Hilger, London, 1971.

\bibitem{candes2012towards}
E.~Candes and C.~Fernandez-Granda.
\newblock Towards a mathematical theory of super-resolution.
\newblock {\em Arxiv preprint arXiv:1203.5871}, 2012.

\bibitem{toraldo52}
G.~Toraldo di~Francia.
\newblock Super-gain antennas and optical resolving power.
\newblock {\em Il Nuovo Cimento (1943-1954)}, 9:426--438, 1952.

\bibitem{toraldo55}
G.~Toraldo di~Francia.
\newblock Resolving power and information.
\newblock {\em Journal of the Optical Society of America}, 45(7):497--501,
  1955.

\bibitem{donoho91}
D.L. Donoho.
\newblock Super-resolution via sparsity constraints.
\newblock Technical Report 285, Department of Statistics, University of
  California, Berkeley, January 1991.

\bibitem{FukGreSunSch08}
K.~Fukumizu, A.~Gretton, X.~Sun, and B.~Sch{\"o}lkopf.
\newblock Kernel measures of conditional dependence.
\newblock In J.C. Platt, D.~Koller, Y.~Singer, and S.~Roweis, editors, {\em
  Advances in Neural Information Processing Systems}, volume~20, pages
  489--496, Cambridge, MA, USA, 09 2008. MIT Press.

\bibitem{gerchberg74}
R.W. Gerchberg.
\newblock Super-resolution through error energy reduction.
\newblock {\em Optica Acta}, 21(9):709--720, 1974.

\bibitem{gerchberg1972practical}
R.W. Gerchberg and W.O. Saxton.
\newblock {A practical algorithm for the determination of phase from image and
  diffraction plane images}.
\newblock {\em Optik (Stuttgart)}, 35:225--246, 1972.

\bibitem{goodman96}
J.W. Goodman.
\newblock {\em Introduction to {F}ourier Optics}.
\newblock McGraw-Hill, second edition, 1996.

\bibitem{GreBorRasSchetal07}
A.~Gretton, K.M. Borgwardt, M.~Rasch, B.~Sch{\"o}lkopf, and A.J. Smola.
\newblock A kernel method for the two-sample-problem.
\newblock In B.~Sch{\"o}lkopf, J.~Platt, and T.~Hofmann, editors, {\em Advances
  in Neural Information Processing Systems}, volume~19, pages 513--520,
  Cambridge, MA, USA, 09 2007. MIT Press.

\bibitem{GreFukTeoSonetal08}
A.~Gretton, K.~Fukumizu, C.H. Teo, L.~Song, B.~Sch{\"o}lkopf, and A.J. Smola.
\newblock A kernel statistical test of independence.
\newblock In J.C. Platt, D.~Koller, Y.~Singer, and S.~Roweis, editors, {\em
  Advances in Neural Information Processing Systems}, volume~20, pages
  585--592, Cambridge, MA, USA, 09 2008. MIT Press.

\bibitem{harris64}
J.L. Harris.
\newblock Diffraction and resolving power.
\newblock {\em Journal of the Opt. Soc. of America}, 54(7):931--946, 1964.

\bibitem{hell1994breaking}
S.W. Hell and J.~Wichmann.
\newblock Breaking the diffraction resolution limit by stimulated emission:
  stimulated-emission-depletion fluorescence microscopy.
\newblock {\em Optics letters}, 19(11):780--782, 1994.

\bibitem{Hirsch11}
M.~Hirsch.
\newblock {\em Blind Deconvolution in Scientific Imaging \& Computational
  Photography}.
\newblock PhD thesis, University of T\"ubingen, 2011.

\bibitem{HofSchSmo08}
T.~Hofmann, B.~Sch{\"o}lkopf, and A.J. Smola.
\newblock Kernel methods in machine learning.
\newblock {\em Annals of Statistics}, 36(3):1171--1220, 06 2008.

\bibitem{irani90}
M.~Irani and S.~Peleg.
\newblock Super resolution from image sequences.
\newblock In {\em Proceedings of the 10th International Conference on Pattern
  Recognition}, pages 115--120, 1990.

\bibitem{jones1986discrete}
M.~Jones.
\newblock The discrete {G}erchberg algorithm.
\newblock {\em Acoustics, Speech and Signal Processing, IEEE Transactions on},
  34(3):624--626, 1986.

\bibitem{papoulis75}
A.~Papoulis.
\newblock A new algorithm in spectral analysis and band-limited extrapolation.
\newblock {\em IEEE Transactions on Circuits and Systems}, CAS-22(9):735--742,
  September 1975.

\bibitem{park2003super}
S.C. Park, M.K. Park, and M.G. Kang.
\newblock Super-resolution image reconstruction: a technical overview.
\newblock {\em Signal Processing Magazine, IEEE}, 20(3):21--36, 2003.

\bibitem{rudin91}
W.~Rudin.
\newblock {\em Functional Analysis}.
\newblock McGraw-Hill, 1991.

\bibitem{rushforth68}
C.K. Rushforth and R.W. Harris.
\newblock Restoration, resolution, and noise.
\newblock {\em Journal of the Optical Society of America}, 58(4):539--545,
  April 1968.

\bibitem{SalTei07}
B.E.A. Saleh, M.C. Teich, and B.E. Saleh.
\newblock {\em Fundamentals of photonics}, volume~22.
\newblock Wiley, 1991.

\bibitem{SchSmo02}
B.~Sch{\"o}lkopf and A.J. Smola.
\newblock {\em Learning with Kernels}.
\newblock MIT Press, Cambridge, MA, USA, 2002.

\bibitem{Sch2008}
B~Sch{\"o}lkopf, BK~Sriperumbudur, A~Gretton, and K~Fukumizu.
\newblock {RKHS} representation of measures applied to homogeneity,
  independence, and {F}ourier optics.
\newblock Technical Report pp.~42--44, OWR 30/2008, Mathematisches
  Forschungsinstitut Oberwolfach, 2008.

\bibitem{sementilli93}
P.J. Sementilli, B.R. Hunt, and M.S. Nadar.
\newblock Analysis of the limit to superresolution.
\newblock {\em Journal of the Optical Society of America A}, 10(11):2265--2276,
  1993.

\bibitem{slepian61}
D.~Slepian, H.O. Pollak, and H.J. Landau.
\newblock Prolate spheroidal wave functions.
\newblock {\em Bell System Technical Journal}, 40:43--84, January 1961.

\bibitem{smola2007hilbert}
A.~Smola, A.~Gretton, L.~Song, and B.~Sch{\"o}lkopf.
\newblock A {H}ilbert space embedding for distributions.
\newblock In {\em Proc. 18th International Conference on Algorithmic Learning
  Theory}, pages 13--31. Springer-Verlag, 2007.

\bibitem{sripermbudur2010}
B.~K. Sriperumbudur, A.~Gretton, K.~Fukumizu, B.~Sch{\"o}lkopf, and
  G.~Lanckriet.
\newblock Hilbert space embeddings and metrics on probability measures.
\newblock {\em Journal of Machine Learning Research}, 11:1517--1561, 2010.

\bibitem{wendland2005scattered}
H.~Wendland.
\newblock {\em Scattered data approximation}.
\newblock Cambridge Univ Pr, 2005.

\bibitem{wolter59b}
H.~Wolter.
\newblock {Verfahren zur beliebig genauen Berechnung einer Originalnachricht
  aus endlich vielen Beobachtungen hinter einem Rechteckbandpa\ss}.
\newblock {\em Archiv der Elektrischen \"Ubertragung (A.E.\"U.)},
  13(9):393--404, 1959.

\end{thebibliography}
}
\end{document}